\documentclass[10pt,conference]{IEEEtran}
\usepackage{graphicx}
\usepackage{amsmath}
\usepackage{amsfonts}
\usepackage{amssymb}

\newtheorem{thm}{Theorem}[section]
\newtheorem{cor}[thm]{Corollary}
\newtheorem{lem}[thm]{Lemma}

\newcommand{\kay}{\eta}
\newcommand{\figref}[1]{\figurename~\ref{#1}}
\newcommand{\secref}[1]{Sec~\ref{#1}}
\newcommand{\appref}[1]{Appendix~\ref{#1}}
\newcommand{\thmref}[1]{Theorem~\ref{#1}}
\newcommand{\lemref}[1]{Lemma~\ref{#1}}
\newcommand{\cn}{\mathcal{CN}}
\newcommand{\rate}{\mathcal{R}}

\newcommand{\ctwo}{I_{\mathcal{C}_{D}}^{'}}
\newcommand{\cone}{I_{\mathcal{C}_{S}}^{'}}

\newcommand{\hbf}{\mathbf{h}}
\newcommand{\Hbf}{\mathbf{H}}
\newcommand{\Ibf}{\mathbf{I}}

\newcommand{\xbf}{\mbox{${\bf X }$} }
\newcommand{\ybf}{\mbox{${\bf Y }$} }

\begin{document}

\title{Cooperative Multiplexing in the Multiple Antenna Half Duplex Relay Channel}

\author{\IEEEauthorblockN{Vinayak Nagpal,
Sameer Pawar, David Tse
and Borivoje Nikoli\'{c}}
\IEEEauthorblockA{EECS Dept. University of California Berkeley, USA.\\
Email: \{vnagpal, spawar, dtse, bora\}@eecs.berkeley.edu}}

\maketitle


\begin{abstract}
Cooperation between terminals has been proposed to improve the reliability and throughput of wireless communication. While recent work has shown that relay cooperation provides increased diversity, increased multiplexing gain over that offered by direct link has largely been unexplored. In this work we show that cooperative multiplexing gain can be achieved by using a half duplex relay. 
We capture relative distances between terminals in the high SNR diversity multiplexing tradeoff (DMT) framework. The DMT performance is then characterized for a network having a single antenna half-duplex relay between a single-antenna source and two-antenna destination. Our results show that the achievable multiplexing gain using cooperation can be greater than that of the direct link and is a function of the relative distance between source and relay compared to the destination. Moreover, for multiplexing gains less than 1, a simple scheme of the relay listening 1/3 of the time and transmitting 2/3 of the time can achieve the 2 by 2 MIMO DMT.

\end{abstract}


\section{Introduction}


There is a growing interest in the design of cooperative schemes that provide diversity and multiplexing gain for communication via wireless relays. Cooperative diversity refers to the additional diversity gain (compared to direct link) offered by cooperation. Similarly if a relay provides additional degrees of freedom (compared to direct link) it is said to provide a cooperative multiplexing gain \cite{4698542}. 

Diversity multiplexing tradeoff (DMT) \cite{1197843} has been widely used to analyze and compare the performance of cooperative schemes. DMT for the half-duplex single relay network has been studied extensively in literature \cite{1362898}\cite{4305423}\cite{sameer}. For the case with single antennas at all terminals the $2\times 1$ MISO DMT bound has recently been shown \cite{sameer} to be achievable. In this paper we study the DMT for the multiple antenna half-duplex relay channel having $m,n$ and $k$ antennas at source, destination and relay respectively. This was studied in \cite{4305423} but results were shown only for the special case $m=n=1$. We calculate the maximum achievable DMT for the $m=1,n=2,k=1$ configuration. In the process we also demonstrate techniques that enable results for general $m,n$ and $k$. $m=1,n=2,k=1$ is the simplest configuration where relay cooperation provides additional multiplexing gain compared to direct link. We show that if source and relay are relatively close to each other, cooperative multiplexing gain is achievable even with half duplex relaying. Note that the full-duplex case has been studied in \cite{4698542}. Moreover, for multiplexing gains less than $1$, if source-relay SNR (measured in dB) is at least two times the source-destination SNR a simple scheme with the relay listening $\frac{1}{3}$ of the time and transmitting $\frac{2}{3}$ of the time can achieve the $2\times2$ MIMO DMT.

\begin{figure}
\centering
\includegraphics[scale=0.3]{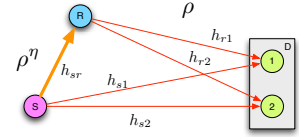}
\caption{Relay channel with $2$ antennas at destination and $S\to R$ proximity gain $\kay$. \label{fig:model}}
\end{figure}

These results lend fresh insight into the fundamental limits of cooperative multiplexing in the half-duplex relay channel. We demonstrate the use of two key techniques that enable our results.
\begin{enumerate}
\item \emph{Distance between terminals:}
In most results it is seen that relative distances between source, relay and destination do not affect DMT performance of the relay channel. Since DMT is calculated at high SNR the path loss and therefore distances are not easily captured in results. 
We overcome this apparent limitation by scaling the average SNR's of the various links differently.

Our approach enriches the DMT framework by adding insights about network geometry. 

\item \emph{MIMO with half duplex antenna:} 
The min-cut capacity bound has been used in \cite{4305423} to calculate an upper bound for DMT performance. Notice in \figref{fig:model} that the $\{S,R\},\{D\}$ cut corresponds to a $2\times 2$ MIMO system with one source antenna that remains active only for a fraction of total communication time ($R$ is half duplex). It was noted  \cite{4305423} that an upper bound for mutual information across such a cut is hard to compute. Due to this, DMT bounds have only been reported for the special case of $m=n=1$.

In \secref{calc} we demonstrate a simple channel decomposition that allows us to compute the cut-set DMT bound for the $m=1,n=2,k=1$ configuration. The technique can be applied towards computing DMT bounds for general $m,n$ and $k$.
Recent results in \cite{4595031} \cite{Avestimehr:EECS-2008-128} show that a simple relaying scheme called ``quantize-map'' can achieve a rate within constant gap of the cut-set capacity.  In \secref{achieve} we discuss this scheme and show that it achieves the cut-set DMT bound.
\end{enumerate}
 
\section{System Model}\label{model}
Consider the system in \figref{fig:model} with source $S$, relay $R$ and destination $D$ having $1,1$ and $2$ antennas respectively. Let $D_{j}, j\in\{1,2\}$ denote the $j$th antenna at $D$. The channel gain for $S \to R$ is $h_{sr}$, gains for $S \to D_{j}$ are $h_{sj}$ and $R \to D_{j}$ are $h_{dj}$. All the channel gains are assumed to be flat fading having i.i.d. $\cn(0,1)$ distribution. We assume quasi-static fading, i.e. once realized, channel gains remain unchanged for the duration of the codeword and change independently between codewords. Noise at all receivers is additive i.i.d. $\cn(0,1)$ and independent of all other variables in the system. Transmit power at $S$ and $R$ is limited by an average power constraint. Since noise power at receiver is normalized to $1$, the transmit power constraint is specified by the average Signal to Noise Ratio (SNR). $R$ is assumed to operate under a half-duplex constraint. For simplicity it is assumed that transmissions at $S$ and $R$ are synchronous at symbol level.

We assume an asymmetrical network geometry. $S$ and $R$ are modeled to be close to each other as compared to $\{S,R\}$ and $D$. $S\to D$ and $R \to D$ are assumed to have the same average SNR denoted by $\rho$. $S\to R$ on the other hand is modeled to have SNR higher than $\rho$ by a factor $\kay$ on dB scale, i.e. the $S\to R$ average SNR is $\rho^{\kay}$. The $S\to R$ channel (cooperation link) thus has $\kay-1$ more degrees of freedom than other channels in the network. We call $\kay$ the proximity gain and assume $\kay\geq1$. 

No  channel state information (CSI) is available at $S$ i.e. only average channel statistics $\rho,\kay$ are known. However, at $D$ all channel realizations $h_{sr}$, $h_{sj},h_{dj}$ are completely known. 

We identify three models for relaying strategy.
\begin{itemize}
\item \emph{Global:} The relay uses knowledge of all instantaneous channel realizations to optimize its strategy.
\item \emph{Local:} The relay can measure $h_{sr}$ and uses only this (local) information.
\item \emph{Blind:} The relay only uses average channel statistics.
\end{itemize}
The \emph{global} strategy is discussed in \secref{results} while \emph{local} and \emph{blind} are discussed in \secref{dynamic}.


\section{Diversity-Multiplexing Tradeoff}\label{results}
\begin{figure}
\centering
\includegraphics[scale=0.5]{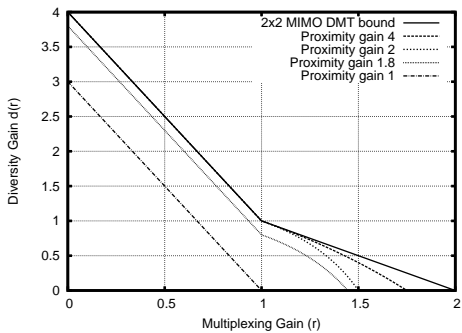}
\caption{$d(r)$ for various values of proximity gain $\kay$.\label{fig:1}}
\end{figure}

\begin{thm}\label{thm:ga}
The maximum achievable DMT for network described in \secref{model} is given by,
\begin{equation} \label{eq:ga}
d(r)=\left\lbrace\begin{array}{cc}
\min\{\kay+2,4\}-3r & 0\leq r\leq1,\kay\geq1 \\
(2\kay-\kay r-1)/(\kay-r) & 1\leq r\leq2-\frac{1}{\kay}, \kay\geq 2 \\
\kay-\frac{1}{2-r} & 1\leq r\leq 2-\frac{1}{\kay}, 1\leq \kay\leq2
\end{array}\right.
\end{equation}
\end{thm}

\begin{cor}
For system model described in \secref{model} the maximum achievable multiplexing gain $r^{*}=\inf_{r\geq0} \{r|d(r)=0\}$ is,
\begin{equation}\label{eq:rmax}
r^{*}=2-\frac{1}{\kay}
\end{equation}
\end{cor}

For a symmetrical geometry with all channels having the same degrees of freedom ($\kay=1$) we get $r^{*}=1$ i.e. cooperation doesn't provide additional maximum multiplexing gain. To enable higher multiplexing gain the $S\to R$ channel (cooperation link) needs to have more degrees of freedom than the $S\to D$ channel (communication link).

Let $d_{2\times 2}(r)$ represent the DMT of the $2\times2$ MIMO channel. For finite $\kay$ it can be seen that $d(r)\leq d_{2\times2}(r)$ with strict inequality over a non-empty region of $r$. This suggests that for distributed antennas the finite capacity of the cooperation channel $(S\to R)$ poses a fundamental limitation on the achievable DMT performance.
It can easily be verified that,
\[
\lim_{\kay\to \infty} d(r) \to d_{2\times2}(r)
\]
\figref{fig:1} shows $d(r)$ for several values of $\kay$.

We prove \thmref{thm:ga} in two steps. In \secref{calc} we show that the cut-set DMT upper bound for network in \secref{model} is given by \eqref{eq:ga}. In \secref{achieve} we show that this bound is achievable.

\subsection{Cut-Set DMT upper bound} \label{calc}
\begin{figure}
\centering
\includegraphics[scale=0.3]{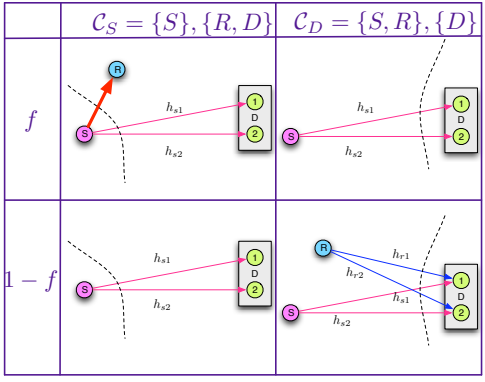}
\caption{Two cuts of network during listen $(f)$ and cooperation phase.$(1-f)$.\label{fig:cuts}}
\end{figure}

Let $f (0\leq f \leq 1)$ denote a listen-transmit schedule for the half duplex relay. $R$ listens for a fraction $f$ (listening phase) of total communication time and transmits  for fraction $(1-f)$ (cooperation phase). The two cuts of the network $\mathcal{C}_{D}=\{S,R\},\{D\}$ and $\mathcal{C}_{S}=\{S\}\{R,D\}$ are shown in \figref{fig:cuts} for these two phases.
In the listening phase let $\mathbf{X}_{S}^{1}$ denote the sequence of symbols transmitted by $S$ while $\mathbf{Y}_{R}$ and $\mathbf{Y}_{D}^{1}$ denote received signals at $R$ and $D$ respectively. Similarly for the cooperation phase $\mathbf{X}_{S}^{2}$ and $\mathbf{X}_{R}$ are the symbol sequences transmitted from $S$ and $R$ while $\mathbf{Y}_{D}^{2}$ is received at $D$. The instantaneous  mutual information across the two cuts can be written as,

{\small
\begin{eqnarray} 
I_{\mathcal{C}_{S}}&=&fI(\mathbf{X}_{S}^{1};\mathbf{Y}_{R},\mathbf{Y}_{D}^{1}|\mathbf{X}_{R})+(1-f)I(\mathbf{X}_{S}^{2};\mathbf{Y}_{D}^{2}|\mathbf{X}_{R}) \label{eq:icut1}\\
I_{\mathcal{C}_{D}}&=&fI(\mathbf{X}_{S}^{1};\mathbf{Y}_{D}^{1})+(1-f)I(\mathbf{X}_{S}^{2},\mathbf{X}_{R};\mathbf{Y}_{D}^{2}) \label{eq:icut2}
\end{eqnarray} }

To maximize these mutual information expressions we need to choose zero-mean complex Gaussian distributions for $\mathbf{X}_{S}^{1},\mathbf{X}_{S}^{2}$ and $\mathbf{X}_{R}$ that have covariance matrices which satisfy their respective average power constraints. Using these distributions we can write mutual information upper bounds $\cone$ and $\ctwo$ for $I_{\mathcal{C}_{S}}$ and $I_{\mathcal{C}_{D}}$ respectively.
{
\begin{eqnarray*}
I_{\mathcal{C}_{S}}\leq I_{\mathcal{C}_{S}}^{'}&=&f\log(1+\rho^{\kay}|h_{sr}|^{2}+\rho||\hbf_{s}||^{2}) \\
&&+(1-f)\log(1+\rho||\hbf_{s}||^{2}) \\
&\approx& f\max\{\log(1+\rho^{\kay}|h_{sr}|^{2}),\log(1+\rho||\hbf_{s}||^{2})\} \\
&& + (1-f)\log(1+\rho||\hbf_{s}||^{2}) \\
I_{\mathcal{C}_{D}}\leq I_{\mathcal{C}_{D}}^{'}&=&f\log(1+\rho||\hbf_{s}||^{2}) \\
&&+(1-f)\log\det(\Ibf+\rho\Hbf \Hbf^{\dagger}) 
\end{eqnarray*} }
where $\hbf_{s}=\left[\begin{array}{c} h_{s1} \\ h_{s2}\end{array}\right], \hbf_{r}=\left[\begin{array}{c} h_{r1} \\ h_{r2}\end{array}\right]$ and $\Hbf=[\begin{array}{cc}\hbf_{s} & \hbf_{r}\end{array}].$
It can be verified that the approximation is tight within one bit.

\begin{figure}
\centering
\includegraphics[scale=0.35]{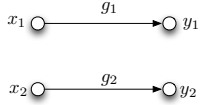}
\caption{Parallel Channel Model for $2\times 2$ MIMO.\label{fig:parchan}}
\end{figure}

Note that the expression for $\ctwo$ is a linear combination of the capacities of two Raleigh fading Gaussian channels having correlated channel matrices $\Hbf_{1}=[\begin{array}{cc}\hbf_{s} & 0 \end{array}]$ and $\Hbf=[\begin{array}{cc}\hbf_{s} & \hbf_{r}\end{array}]$. Outage probability for the MIMO channel was calculated in \cite{1197843} using eigenvalue decomposition of the channel matrix. Following the same technique there would require computing the joint eigenvalue distributions for the two correlated hermitian matrices, $\Hbf_{1}\Hbf_{1}^{\dagger}$ and $\Hbf \Hbf^{\dagger}$. It was noted in \cite{4305423} that this is hard to compute. We propose an easier decomposition to solve this problem. The second term in $\ctwo$ is the capacity of a $2\times2$ MIMO channel which can be represented by two parallel Gaussian channels having gains $g_{1}$ and $g_{2}$ shown in \figref{fig:parchan}.
The channel can be written as, {
\[
\left[\begin{array}{c}y_{1} \\ y_{2}\end{array}\right]=\sqrt{\rho}\left[\begin{array}{cc}g_{1}& 0 \\ 0 & g_{2}\end{array}\right]\left[\begin{array}{c}x_{1} \\ x_{2}\end{array}\right]+\left[\begin{array}{c}w_{1} \\ w_{2}\end{array}\right]
\] }
where $E[|x_{i}|^{2}]=1$, $w_{1},w_{2}\sim\cn(0,1)$. The capacity for $x_{i}\to y_{i}$ is given by $\log(1+\rho g_{i}^{2})$.
It was shown in \cite{679134}\cite{1197843} that a D-BLAST transmission scheme
 with a MMSE successive interference cancellation receiver achieves the mutual information of the MIMO channel. For this scheme $g_{1}^{2}$ and $g_{2}^{2}$ can be calculated to be,
{ \begin{eqnarray*}
g_{2}^{2}&=&||\mathbf{h}_{r\perp s}||^{2}+\frac{||\mathbf{h}_{r\parallel s}||^{2}}{1+\rho||\mathbf{h}_{s}||^{2}}\\
g_{1}^{2}&=&||\mathbf{h}_{s}||^{2}
\end{eqnarray*} }
where $\mathbf{h}_{r\perp s}$ and $\mathbf{h}_{r\parallel s}$ respectively denote the perpendicular and parallel components of $\mathbf{h}_{r}$ with respect to $\mathbf{h}_{s}$. 

Note that while $g_{1}^{2}$ and $g_{2}^{2}$ are correlated, $\mathbf{h}_{s},\mathbf{h}_{r\perp s}$ and $\mathbf{h}_{r\parallel s}$ are mutually independent. The correlation between $g_{1}^{2}$ and $g_{2}^{2}$ can therefore be explicitly calculated. The destination decodes $\mathbf{X}_{R}$ in the presence of interference from $\mathbf{X}_{S}^{2}$. It then cancels $\mathbf{X}_{R}$ from its received signal before decoding $\mathbf{X}_{S}^{2}$. Therefore $S$ effectively sees an interference free channel (with gain $g_{1}$) to $D$ during both listen and cooperation phases.
{
\begin{eqnarray*}
\ctwo&=&f\log(1+\rho g_{1}^{2}) \\
&&+(1-f)[\log(1+\rho g_{1}^{2})+\log(1+\rho g_{2}^{2})] \\
&=& \log(1+\rho g_{1}^{2})+(1-f)\log(1+\rho g_{2}^{2})
\end{eqnarray*} }

Let $\alpha_{sr},\alpha_{1}$ and $\alpha_{2}$ represent channel realizations via the following variable transformations. 
{
\begin{eqnarray*}
\alpha_{sr}&=&\lim_{\rho\to\infty}\frac{\log(1+\rho^{\kay} |h_{sr}|^{2})}{\log{\rho}} \\
\alpha_{1}&=&\lim_{\rho\to\infty}\frac{\log(1+\rho g_{1}^{2})}{\log{\rho}} \\
\alpha_{2}&=&\lim_{\rho\to\infty}\frac{\log(1+\rho g_{2}^{2})}{\log{\rho}}\end{eqnarray*} } 
This gives us simplified expressions for mutual information upper bounds.
{ 
\begin{eqnarray}
\frac{\cone}{\log \rho}&=&\alpha_{1}+f(\alpha_{sr}-\alpha_{1})^{+} \label{eq:cone}\\
\frac{\ctwo}{\log \rho}&=&\alpha_{1}+(1-f)\alpha_{2} \label{eq:ctwo}
\end{eqnarray} }

To achieve desired multiplexing gain $r$ at high SNR ($\rho \to \infty$) the network must achieve a rate $\rate=r\log\rho$. The network is in outage if, $\min\{\cone,\ctwo\}\leq r\log\rho$. For a given $r$  and schedule $f$ we can define the outage region $\mathcal{O}(r,f)$ over channel realizations $\vec{\alpha}=(\alpha_{1},\alpha_{2},\alpha_{sr})$.
{ \begin{equation}
\mathcal{O}(r,f)=\{\vec{\alpha}|\frac{\min\{\cone,\ctwo\}}{\log\rho}\leq r\}
\end{equation} }

The outage probability $\mathcal{P}_{out}$ is, {
\[
\mathcal{P}_{out}=\int_{\vec{\alpha}\in\mathcal{O}(r,f)} f_{\vec{\alpha}}(\alpha_{1},\alpha_{2},\alpha_{sr})
\] }
where $f_{\vec{\alpha}_{1}}(\alpha_{1},\alpha_{2},\alpha_{sr})$ is the joint distribution of $(\alpha_{1},\alpha_{2},\alpha_{sr})$.

\begin{lem}{Proof see \appref{appa}}{
\[
f_{\vec{\alpha}}(\alpha_{1},\alpha_{2},\alpha_{sr})\doteq \rho^{-s(\vec{\alpha})}
\] } where $0\leq \alpha_{1},\alpha_{2}\leq 1$, $0 \leq \alpha_{sr}\leq \kay$ and 
\begin{equation}
s(\vec{\alpha})=\left\lbrace\begin{array}{cc}\kay+4-3\alpha_{1}-2\alpha_{2}-\alpha_{sr} & \alpha_{1}+\alpha_{2}\leq 1 \\ \kay+3-2\alpha_{1}-\alpha_{2}-\alpha_{sr} & \alpha_{1}+\alpha_{2}>1\end{array}\right.
\end{equation}  \label{lem1}
\end{lem}

For a given listen-transmit schedule $f$ the cut-set DMT bound is therefore given by,
\begin{equation}
d(r,f)=\min_{\vec{\alpha}\in \mathcal{O}(r,f)} s(\vec{\alpha}) \label{eq:drt}
\end{equation}
To get the DMT upper bound we can optimize over all listen-transmit schedules, 
\begin{equation}
d(r)=\min_{\vec{\alpha}\in \mathcal{O}(r)} \max_{f} s(\vec{\alpha})
\end{equation}
Note that this optimization is performed on a per realization basis, i.e. the optimal $f$ depends on all channel realizations $\alpha_{sr},\alpha_{1}$ and $\alpha_{2}$.  Therefore this corresponds to the \emph{global} strategy discussed in \secref{model}. 

It is easy to see that the globally optimal schedule $f_{glob}$ is one which sets $\cone=\ctwo$. 
\[
f_{glob}=\frac{\alpha_{2}}{(\alpha_{sr}-\alpha_{1})^{+}+\alpha_{2}}
\]
This leads to the solution for $d(r)$ given in \eqref{eq:ga}.

\section{Achievability: Relaying Scheme} \label{achieve}
 The ``quantize-map" relaying scheme proposed in \cite{4595031} and \cite{Avestimehr:EECS-2008-128} has been shown to
be DMT optimal for the single antenna relay channel \cite{sameer}. We show that ``quantize-map'' adapts naturally to the network described in \secref{model} and with some modification achieves the cut-set DMT bound. For the sake of completeness we include a short description of the scheme.
\subsection{Description of scheme}\label{scheme}
$S$ has a sequence of messages $w_n
\in\{1,2,\ldots,2^{T\rate}\}$, $n=1,2,\ldots$ to be transmitted. At
both source $S$ and relay we create random Gaussian codebooks.
$S$ randomly maps each message to one of its Gaussian codewords
and transmits it using $T$ symbol times giving an overall
transmission rate of $\rate$. Due to the half-duplex nature of the relay, it
must operate using listen-transmit cycles. Relay listens to the first $fT$
 time symbols of each block i.e. $\mathbf{X}_{S}^{1}$. It quantizes $\mathbf{Y}_{R}$ to $\hat{\mathbf{Y}}_R$ and then randomly maps it
into a Gaussian codeword $\xbf_R$ using a random mapping
function $f_R(\hat{ \ybf}_R)$. It transmits this codeword during the next
$(1-f)T$ symbol times. Given the
knowledge of all the encoding functions and signals
received, $D$ attempts to decode the
message sent by $S$.

\subsection{DMT of Quantize-Map}
By Theorem 7.4.1 in \cite{Avestimehr:EECS-2008-128}, for any fixed
listen-transmit schedule $f$, the quantize-map
relaying scheme, uniformly over
all channel realizations achieves a rate within a constant gap to
the cut-set upper bound $\min\{\cone,\ctwo\}$ for that particular $f$. The random Gaussian code-book generated at source is independent of $f$. Also the code-book generated at relay depends on $f$ only to determine the length of each codeword $(1-f)T$. 

The relay can generate a larger code-book with each codeword of
length $T$. If the relay now chooses a listen-transmit schedule $f$, it can use the   first $(1-f)T$ symbols of the codeword to compose
$\xbf_R$. The destination always knows the schedule $f$ and hence can adapt its decoder accordingly. This construction allows us to claim that ``quantize-map" achieves a rate within a constant gap of $\min\{\cone,\ctwo\}$ uniformly for each dynamic choice of $f$ i.e. 
{
\begin{equation} \label{eq:QMRate}
\min\{\cone,\ctwo\} -\kappa \leq
R_{\text{quantize-map}}(h_{sr},\hbf_{s},\hbf_{r},\rho,f)
\end{equation}}

The constant $\kappa$ in the above equation does not depend on the
channel gains and $\mbox{SNR}$. At the order of DMT which assumes high SNR $(\rho\to \infty)$ the effect of $\kappa$ becomes negligible and hence we have the following theorem for achievability.

\begin{thm}\label{thm:achieve}
For dynamic listen-transmit schedules, the modified quantize-map
relaying scheme as described above achieves the diversity
multiplexing tradeoff of $\min\{\cone,\ctwo\}$, where
$\cone$ and $\ctwo$ are given by \eqref{eq:cone}\eqref{eq:ctwo}.
\end{thm}

\section{Achievability: Listen-Transmit Schedule}\label{dynamic}
In \secref{calc} the cut-set DMT upper bound was calculated for the \emph{globally} optimal listen-transmit schedule $f_{glob}$. However in a practical communication scenario global knowledge of instantaneous channel realizations may not be available at the relay. To account for this we defined the \emph{local} and \emph{blind} relaying strategies in \secref{model}.
In this section we refine \thmref{thm:ga} to calculate DMT bounds for \emph{local} and \emph{blind} schedules. 

\subsection{Blind Scheduling}
\begin{thm}\label{thm:lowrate}
For the low rate region i.e. $r\leq1$, the \emph{blind} scheduling strategy is DMT optimal. Additionally for $\kay\geq2$ the \emph{blind} strategy achieves the  $2\times2$ MIMO DMT bound for $r\leq1$. The optimal blind schedule for this region is $f_{blind}=\frac{1}{3}$
\begin{equation}
\begin{array}{cc}
d_{blind}(r)=\min\{\kay+2,4\}-3r & r\leq 1 
\end{array}
\end{equation}
\end{thm}

\begin{figure}
\centering
\includegraphics[scale=0.7]{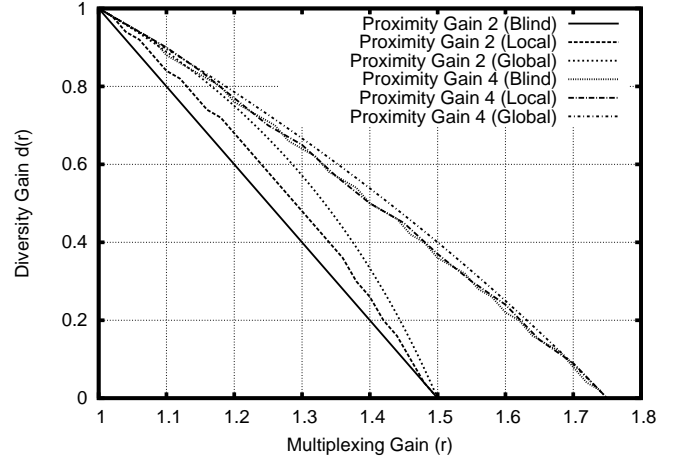}
\caption{$d(r)$ $d_{local}(r)$ and $d_{blind}(r)$ comparison for $r\geq 1$.\label{fig:2}}
\end{figure}

From \eqref{eq:drt} the DMT bound for blind scheduling can be written as,
\begin{equation}
d_{blind}(r)= \max_{f} \min_{\vec{\alpha}\in \mathcal{O}(r,f)} s(\vec{\alpha}) \label{eq:blind}
\end{equation}
i.e. $f$ is optimized without knowledge of channel realizations $\vec{\alpha}$. Solving this optimization for $r\leq1$ yields \thmref{thm:lowrate}.
This suggests that as long as cooperative multiplexing is not necessary i.e. desired rate $\rate=r\log(\rho)$ is such that $r\leq1$, static scheduling at relay is sufficient to achieve the DMT upper bound. $f_{blind}=\frac{1}{3}$ turns out to be the optimal listen-transmit schedule for this region.

For the high rate region $(r>1)$, the analytical solution for \eqref{eq:blind} is tedious to obtain. The optimization is convex and can be solved numerically. \figref{fig:2} shows a comparison between $d(r)$ and $d_{blind}(r)$ for $r\geq1$. It can be seen that for cooperative multiplexing $(r\geq1)$ static scheduling is insufficient to achieve DMT upper bound.

\subsection{Local Scheduling}
Similarly, the DMT bound for local scheduling can be expressed as an optimization problem from \eqref{eq:drt}.
\begin{equation}
d_{local}(r)=\min_{\alpha_{sr}} \max_{f} \min_{\alpha_{1},\alpha_{2}\in\mathcal{O}(r,f)} s(\vec{\alpha})
\label{eq:local}
\end{equation}
$f_{local}$ can be optimized using knowledge of $\alpha_{sr}$ only. The DMT performance of local scheduling must be at-least as good as blind scheduling, therefore by \thmref{thm:lowrate} for $r\leq 1$ $d_{local}(r)=d_{blind}(r)=d(r)$.

Numerical solution to \eqref{eq:local} for the high rate $r\geq1$ region is shown in \figref{fig:2}. It can be seen that local scheduling performs better than blind, but for higher $\kay$ this advantage diminishes.  

\section{Acknowledgements}
The authors wish to acknowledge the contributions of the students, faculty and sponsors of the Berkeley Wireless Research Center and the National Science Foundation Infrastructure Grant No. 0403427.

\appendices

\section{Proof of \lemref{lem1}}\label{appa}
\subsection{Marginal Distribution of $\alpha_{sr}$}
$f_{\alpha_{sr}}(\alpha)$ is calculated as,
\begin{eqnarray*}
\mathcal{P}[\alpha_{sr}<\alpha]&=&\lim_{\rho\to\infty}\mathcal{P}[|h_{sr}|^{2}<\rho^{\alpha_{sr}-\kay}] \\
f_{\alpha_{sr}}(\alpha)&\doteq&\rho^{\kay-\alpha} (0 \leq \alpha_{sr}\leq \kay)
\end{eqnarray*}
\subsection{Joint Distribution of $\alpha_{1}$ and $\alpha_{2}$}
Note that $g_{1}^{2}$ has a $\chi^{2}_{4}$ distribution, the marginal distribution of $\alpha_{1}$ is given by,
\begin{align*}
f_{\alpha_{1}}(\alpha_{1})&=f_{g_{1}^{2}}(\rho^{-(1-\alpha_{1})}) \frac{dg_{1}^{2}}{d\alpha_{1}} \\
&\doteq \rho^{-2(1-\alpha_{1})} (0\leq\alpha_{1}\leq1)
\end{align*} 
Now for $(0\leq\alpha_{1},\alpha_{2}\leq1)$ their joint CDF can be written as,
\begin{eqnarray*}
F_{\alpha_{1},\alpha_{2}}(\alpha_{1},\alpha_{2})&=&\mathcal{P}[g_{1}^{2}\leq \rho^{\alpha_{1}-1},g_{2}^{2}\leq \rho^{\alpha_{2}-1}] \\
&=&\int_{x=0}^{\alpha_{1}} \mathcal{P}[g_{1}^{2}= \rho^{x-1},g_{2}^{2}\leq \rho^{\alpha_{2}-1}] dx \\
&\doteq& \int_{x = 0}^{\alpha_1}\rho^{-2(1-x)} \\ && \mathcal{P}(|h_{s\perp r}|^2 +
\frac{|h_{s\parallel r} |^2}{1 + \rho^{x}} \leq \rho^{\alpha_2 - 1})dx \\
&\doteq& \int_{x = 0}^{\alpha_1}\rho^{-2(1-x)} \\ && \int_{y=0}^{\alpha_2
-1}\rho^{y}\mathcal{P}(|h_{s\parallel r} |^2\dot\leq \rho^{x+\alpha_2 - 1})\ dy\ dx\\
\end{eqnarray*} 
\subsubsection{Case $\alpha_{1}+\alpha_{2}\leq 1$}
\begin{eqnarray*}
F_{{\alpha_1,\alpha_2}}(\alpha_1,\alpha_2) &\doteq& \int_{x =
0}^{\alpha_1}\rho^{-2(1-x)}\int_{y=0}^{\alpha_2
-1}\rho^{x+y+\alpha_2 - 1}\ dy\ dx \\
&\doteq& \int_{x = 0}^{\alpha_1}\rho^{3x+2\alpha_2 - 4}\ dx\\
&\doteq& \rho^{3\alpha_1+2\alpha_2 - 4}\\
f_{\alpha_1,\alpha_2}(\alpha_{1},\alpha_{2}) 
&\doteq& \rho^{3\alpha_1+2\alpha_2 - 4}
\end{eqnarray*}
\subsubsection{Case $\alpha_{1}+\alpha_{2} > 1$}
\begin{eqnarray*}
F_{{\alpha_1,\alpha_2}}(\alpha_1,\alpha_2) &\doteq& \int_{x = 0}^{1-\alpha_2}\rho^{3x+2\alpha_2 - 4}\ dx 
\\&& +\int_{x =1-\alpha_2}^{\alpha_1}\rho^{2x + \alpha_2 -3}\ dx\\
&\doteq& \rho^{2\alpha_2 + \alpha_1 -3} \\
f_{\alpha_1,\alpha_2} 
&\doteq& \rho^{2\alpha_1 + \alpha_2 -3}
\end{eqnarray*}
Since $\alpha_{sr}$ is independent of $\alpha_{1},\alpha_{2}$ we get \lemref{lem1}.

\bibliographystyle{IEEEtran} 
\bibliography{IEEEabrv,isit09} 
\end{document}